\documentclass[%
 reprint,
 amsmath,amssymb,
 aps,
 physrev,
]{revtex4-2}

\usepackage{graphicx}% Include figure files
\usepackage{dcolumn}% Align table columns on decimal point
\usepackage{bm}% bold math
\usepackage{xcolor}
\usepackage{hyperref}% add hypertext capabilities
\hypersetup{colorlinks=true, citecolor=blue, urlcolor=blue, linkcolor=blue}

\newcommand{\degree}{$^\circ$}
\newcommand{\vud}{$\lvert V_{ud}\rvert$}

\newcommand{\icarus}{Icarus}
\newcommand{\jgr}{J. Geophys. Res.}
\newcommand{\planss}{Planet. Space Sci.}
\newcommand{\ssr}{Space Sci. Rev.}

\begin{document}

\title{Space-Based Measurement of the Neutron Lifetime using Data from the Neutron Spectrometer on NASA's MESSENGER Mission}

\author{Jack T. Wilson}
 \email{Jack.Wilson@jhuapl.edu}
\author{David J. Lawrence}%
\author{Patrick N. Peplowski}%
 % \email{Second.Author@institution.edu}
\affiliation{%
 The Johns Hopkins Applied Physics Laboratory,\\
 11101 Johns Hopkins Road,\\
 Laurel, Md. 20723, USA.
}%

\author{Vincent R. Eke}
\author{Jacob A. Kegerreis}
\affiliation{
 Institute for Computational Cosmology,\\ 
 Durham University, South Road, \\
 Durham DH1 3LE, UK.
}%

\date{\today}

\begin{abstract}
We establish the feasibility of measuring the neutron lifetime via an alternative, space-based class of methods, which use neutrons generated by galactic cosmic ray spallation of planets’ surfaces and atmospheres.  Free neutrons decay via the weak interaction with a mean lifetime of around 880~s. This lifetime constrains the unitarity of the CKM matrix and is a key parameter for studies of Big-Bang nucleosynthesis. However, current laboratory measurements, using two independent approaches, differ by over 4$\sigma$. Using data acquired in 2007 and 2008 during flybys of Venus and Mercury by NASA's MESSENGER spacecraft, which was not designed to make this measurement, we estimate the neutron lifetime to be $780\pm60_\textrm{stat}\pm70_\textrm{syst}$~s, thereby demonstrating the viability of this new approach.
\end{abstract}

% \pacs{Valid PACS appear here}% PACS, the Physics and Astronomy
                             % Classification Scheme.
%\keywords{Suggested keywords}%Use showkeys class option if keyword
                              %display desired
\maketitle

\section{\label{sec:Introduction}Introduction}
Measurement of the neutron lifetime $\tau_n$ via space-based observation was first proposed in 1990 by \citet{Feldman1990}.  However, no mission designed to measure $\tau_n$ has flown and no previous successful measurements of $\tau_n$ from space have been reported. We present a measurement of $\tau_n$ using data taken by NASA's MESSENGER spacecraft \citep{Solomon2007} during its flybys of Venus and Mercury \citep{McNutt2008}. These data were taken with the aim of characterizing the compositions of Mercury's surface, so this measurement technique is not optimized and there are systematics present that could be avoided with a dedicated mission. However, the measurement's success demonstrates the possibility of measuring $\tau_n$ from space. The new technique exploits the fact that $\tau_n$ affects both the relative count rates at Venus and Mercury and the rate at which the neutron flux decreases with distance from the top of Venus' atmosphere. During MESSENGER's flyby of Venus the mean time of flight for a detected neutron varied between 80~s and 640~s.

$\tau_n$ is an important parameter for cosmology, particle and nuclear physics \citep{Wietfeldt2011,Dubbers2011}.  In particular, uncertainties on $\tau_n$ currently dominate those on estimates of the primordial helium abundance \citep{Mathews2005}.  Additionally, the unitarity of the Cabibbo--Kobayashi--Maskawa (CKM) matrix is one of the most important low-energy tests of the standard model. Along with the neutron axial-current coupling constant, $\tau_n$ can be used to determine the CKM matrix element \vud{}. Although the most precise determination of \vud{} is currently obtained from measurement of the half-lives of superallowed $0^+ \ensuremath{\rightarrow} 0^+$ nuclear $\beta$ decays \citep{Hardy2015}, using $\tau_n$ provides a theoretically cleaner  measurement.  This is of particular importance, given that recent updates to the universal radiative correction for $\beta$ decay and the resulting estimate of \vud{} place the CKM matrix in some tension with unitarity \citep{Seng2018}.

\begin{figure}
\includegraphics[width=\columnwidth]{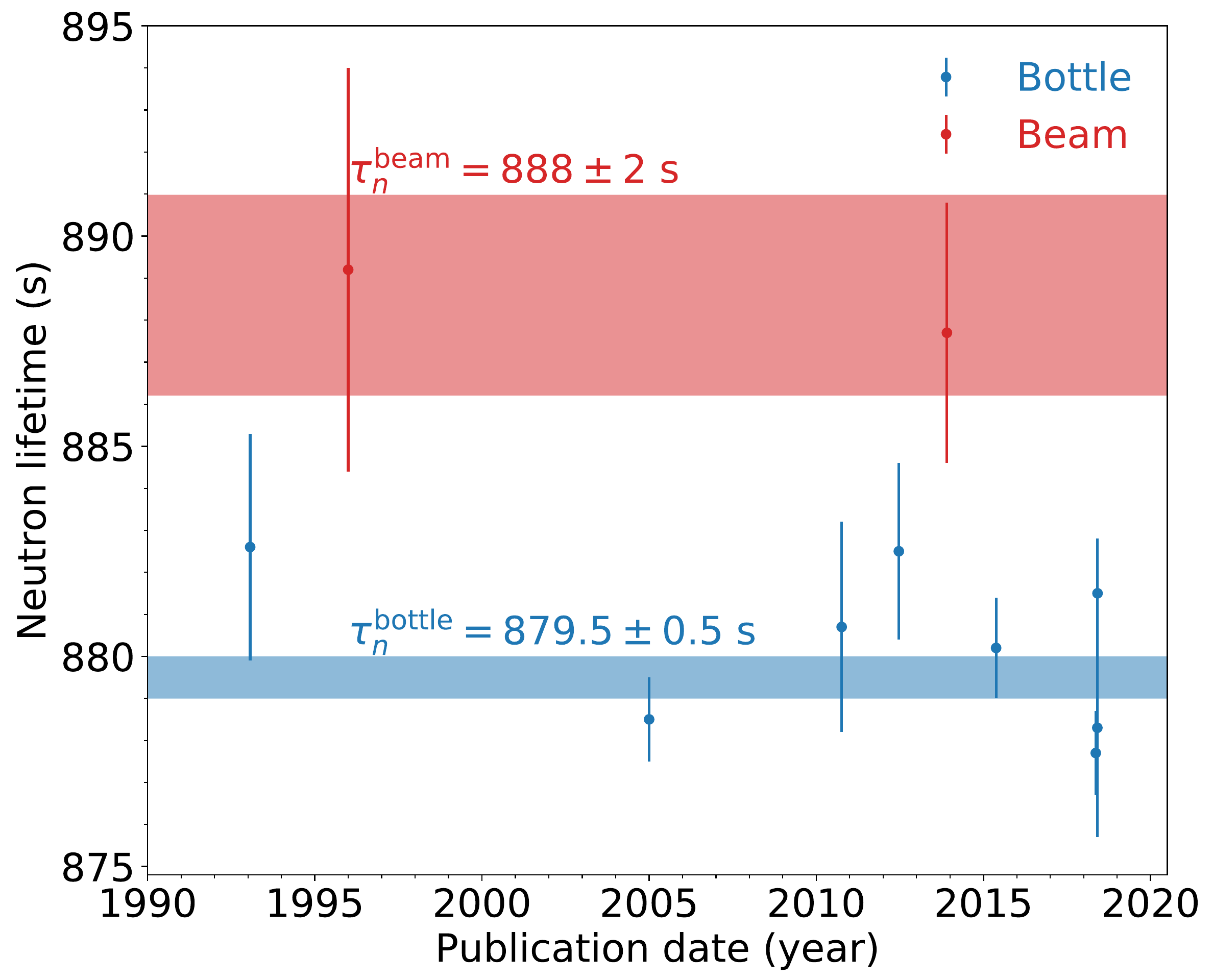}
\caption{\label{fig:history} PDG and more recent measurement of $\tau_n$ \citep{PDG,Pattie2018,Serebrov2018,Ezhov2018}. The shaded regions represent the standard error on the uncertainty-weighted mean lifetime in each class.}
\end{figure}

There are two competing values for $\tau_n$ based on the results of two different classes of experiments.  The `bottle' methods count the number of neutrons $N$ that remain within a mechanical, gravitational, and/or magnetic trap as a function of time, with  $\tau_n$ determined from the exponential decay function ${N(t)=N(0)e^{-t/\tau_n}}$. Thus, these experiments are sensitive to the decay of neutrons by any decay channel.  Alternatively, the `beam' methods measure the protons or electrons resulting from $\beta$ decay, with $\tau_n$ determined from the differential form of the exponential function ${dN/dt = -N/\tau_n}$. The average beam measurements ${\tau_n^\textrm{beam} = 888\pm2}$~s differ by about 4$\sigma$ from the more precise ultra cold trapped neutron average ${\tau_n^\textrm{bottle} = 879.5\pm0.5}$~s (Fig.~\ref{fig:history}). The current data used in the Particle Data Group (PDG) estimate of $\tau_n$ are shown in Fig.~\ref{fig:history}.  Individual experiments report uncertainties smaller than 1~s.  However, given the 9~s (or 4$\sigma$) disagreement between the two classes of measurement, the small error on individual measurements is not representative of our uncertainty on $\tau_n$.  Given the direction of the disagreement, a physical explanation of the neutron lifetime problem can be constructed by positing a decay outside of the standard model into the dark sector \citep[e.g.,][]{Fornal2018,Tan2019}. This again shows the possibility of testing the standard model at low energy using cold neutrons.

% \subsection{\label{ssec:nuclear}Planetary nuclear spectroscopy}
The ability to make a space-based measurement of $\tau_n$ is made possible by the fact that planetary atmospheres and---for airless bodies---solid surfaces are constantly bombarded by galactic cosmic rays (GCRs).  These energetic particles, which are mostly high-energy protons, collide with atomic nuclei leading to spallation reactions in which large numbers of high-energy neutrons are produced.  These neutrons undergo further collisions with nuclei and have their energy moderated downwards.  A fraction of the neutrons undergo a sufficiently large number of collisions that they reach thermal equilibrium with the atmosphere or solid surface.  The energy distribution of neutrons that ultimately escape from a planet into space is characteristic of the planet's elemental composition on depth scales of order the neutron mean free path \citep{Lingenfelter1961}.  As measuring planets' compositions often forms a major goal of NASA's planetary missions, several neutron spectrometers have been flown into space to achieve this aim \citep{Feldman2004,Boynton2004,Goldsten2007,Prettyman2011}.

% \subsection{\label{ssec:MESSENGER}MESSENGER mission to Mercury}
Here, we make use of data taken by NASA's MESSENGER spacecraft to demonstrate the feasibility of space-based measurement of $\tau_n$.  The MESSENGER neutron spectrometer (NS) was designed to measure Mercury's surface composition with special emphasis placed on testing the hypothesis that the radar bright regions at Mercury's poles are a consequence of the presence of water-ice in the permanently shadowed craters \citep{Solomon2001}. MESSENGER's neutron detector consisted of a $10^3$~cm$^3$ cube of borated plastic scintillator sandwiched between two 4~mm thick, 100~cm$^2$ Li-glass plates \citep{Goldsten2007}.  These Li-glass detectors were sensitive to neutrons with energies in the thermal regime, via the neutron capture reaction ${^6\textrm{Li} + n \ensuremath{\rightarrow} {}^3\textrm{H} + {}^4}$He, with decreasing sensitivity into the epithermal range \citep{Goldsten2007}.

Before MESSENGER achieved orbit around Mercury, it carried out a set of flybys at the Earth, Venus, and Mercury using gravity assist maneuvers to alter its trajectory.  During the second flyby of Venus the NS was turned on and collected data \citep{McNutt2008}.  The data used in this study were taken during this Venus encounter and during the first flyby of Mercury \citep{Lawrence2010}.

\section{\label{sec:newMethod}Using MESSENGER data to measure $\tau_n$}
Earlier proposals for space-based measurement of $\tau_n$ would measure the upward-directed flux of neutrons either using a set of neutron detectors arranged in a ring and separated from one another by neutron absorbing material \citep{Nieto2008} or by changing the orientation of a plate detector \citep{Feldman1990}.  From this measured, upward-directed flux that in the downward direction could be calculated for a particular lifetime, as the masses of the planets are well known.  Comparison of this inferred lifetime with the flux measured by upward facing detectors would enable $\tau_n$ to be measured, with appropriate correction for the non-uniformity of the planet's surface and/or atmospheric composition.

As MESSENGER has only two opposing neutron detectors and the frequently changing spacecraft orientation was determined by thermal constraints and the requirements of the spacecraft's imagers, analytic propagation of upward to downward going neutrons is not possible. MESSENGER did, however, sample the neutron fluxes at Venus and Mercury at all altitudes above closest approach (339~km for Venus, 205~km for Mercury) during the flybys. 

In our new approach $\tau_n$ is determined by comparing the output of a set of models based on different lifetimes to the data measured at Mercury and Venus. $\tau_n$, along with surface or atmospheric composition and the planet's mass, determines the rate at which the neutron flux decreases with increasing distance from the planets.

The composition of Venus' atmosphere, from where detected neutrons originate, is both simple and relatively uniform. The atmosphere is principally comprised of only two components: CO$_2$ makes up approximately 96\% by volume with the remaining part composed almost entirely of N$_2$ \citep{vonZahn1983}.  Since nitrogen is an effective neutron absorber via the  $^{14}\textrm{N}+n\ensuremath{\rightarrow}{}^{15}$N and $^{14}\textrm{N}+n\ensuremath{\rightarrow}{}^{14}\textrm{C}+p$ reactions, its abundance has a strong effect on the Venus-originating thermal neutron flux \citep{Lingenfelter1962} that we use to measure $\tau_n$. Venus' homopause is at 120~km \citep{Mahieux2017}, which is above the altitude at which the thermal neutrons originate (60--80~km \citep{Peplowski2019}). Beneath the homopause the atmosphere is uniform as different species are homogeneously mixed by eddy diffusion and turbulent mixing (though recent evidence implies the existence of a discontinuity at an altitude of 50~km \citep{Peplowski2019}). As the neutron flux is originating beneath the homopause we can consider Venus' atmosphere to be compositionally uniform.  Atmospheric temperature is also important as the detected thermal neutrons are in thermal equilibrium with the atmosphere, and so have their energy distribution determined by its temperature.  The temperature, at the altitudes from which neutrons are sourced, varies little over time with large variations not seen beneath 100~km \citep{Limaye2017}. Latitudinal variation in temperature is not seen between 30\degree~S and 30\degree~N \citep{Seiff1985}, where the closest approach of the flyby took place, and is less than 30\% globally. For the Venus flyby a set of models with different neutron lifetimes and atmospheric N$_2$ abundances were generated.

In addition to Venus' atmospheric uniformity, the planet's relatively large mass is advantageous when measuring $\tau_n$. Thermal neutrons have an energy less than 1~eV.  Since Venus' gravitational binding energy is 0.56~eV,  $\tau_n$ has a substantial effect on the neutron flux at all altitudes where a thermal neutron flux is detectable. The basis of this measurement technique is a comparison of the measured Li-glass-derived data with models of the count rate constructed assuming different values of $\tau_n$. 

While the compositional variations on Mercury’s surface complicate the use of the Mercury data for a neutron lifetime measurement, the data were useful for model normalization. Unlike Venus' atmosphere, Mercury's surface is not spatially uniform and contains a large number of elements present at levels high enough to affect the planet-originating neutron flux.  Many of these elements were mapped by MESSENGER during its 4-year orbital mission \citep[e.g.][]{Peplowski2015,Lawrence2017,VanderKaaden2017,Peplowski2019JGR,Wilson2019}; consequently we have a good understanding of Mercury's composition on large scales. For the Mercury flyby, a set of models with different neutron lifetimes but constant composition were generated. Constant composition was required as the parameter space of potential surface compositions is too large to explore fully.

\section{\label{sec:Methods}Modeling and Data Reduction}
MESSENGER's flyby of Venus with its instruments turned on occurred at 23:08 Coordinated Universal Time (UCT) on 5~June~2007 \citep{McNutt2008}. The first flyby of Mercury, the data from which were used to derive the model normalization factors, occurred on 14 January 2008. At both planets our analysis is restricted to data acquired at altitudes below $10^4$~km.  This altitude limit was chosen because above this height the measured count rate was not significantly different from the background, therefore including any higher-altitude data has no effect on the inferred value of $\tau_n$.

% \subsection{\label{ssec:Data reduction}}
During the flybys, each Li-glass detector recorded a 64-channel energy-deposition spectrum every two seconds.  On the ground, these two-second spectra were combined into 20-second spectra for a total of 133 20-second observations at Venus and 78 at Mercury. The measured spectra include the 4.78~MeV energy deposition peak from the $^6\textrm{Li}(n,\alpha){}^3\textrm{H}$ neutron-capture reaction, which is the primary signal of interest, as well as a continuum background due to the interaction of GCRs and high-energy, planet-originating neutrons with the spacecraft and detector \citep{Lawrence2010}.  The conversion of recorded spectra to count rates involved removing this background and then summing over the channels that measure the neutron absorption peak. 

\begin{figure}
 \includegraphics[width=\columnwidth]{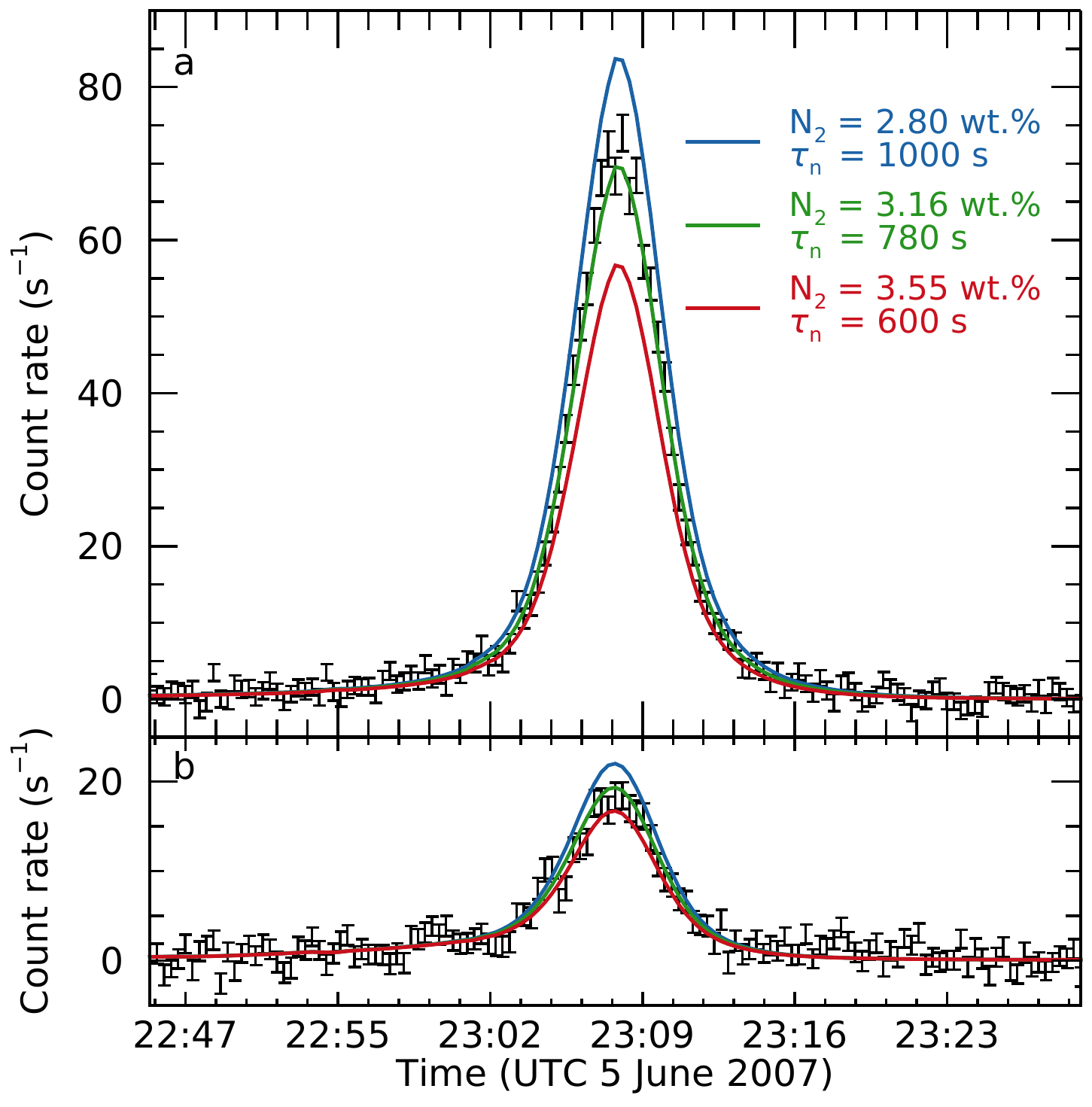}
\caption{\label{fig:cts} Modeled and measured data taken by MESSENGER's Neutron Spectrometer when the spacecraft was within $10^4$~km of Venus' surface for the Li-glass detectors with their surface normals approximately (a) aligned with and (b) opposite the direction of spacecraft motion.}
\end{figure}

The background spectrum was estimated by summing all of the spectra taken during the flybys when the spacecraft altitude was greater than $10^4$~km. Background-subtracted spectra were then produced by subtracting a scaled version of this high-altitude background from each spectrum in the time series, with the scaling factor chosen such that the total counts in spectra away from the neutron absorption peak were the same in the low-altitude observation and scaled background.  Finally, to calculate the neutron count rate, the background-subtracted spectra were summed over the channels containing the neutron-absorption peak before being divided by the observation period.  Uncertainties are those resulting from the Poisson statistics of the observed spectra. The data from both of the Li-glass detectors are shown in Fig.~\ref{fig:cts}.

% \subsection{\label{ssec:Modelling}}

The modeled count rates were determined using three separate calculations.  First, we used the particle transport code MCNPX \citep{Pelowitz2005} to model the neutron flux escaping the planets' surface or atmosphere. For Venus the MCNPX geometry included the planet's solid surface along with a 100-km-thick atmosphere that was split into 50 2-km-thick layers. Each layer had uniform composition but altitude-dependent variations in the temperature and density that reflect the actual variation in Venus' atmosphere \citep{Kliore1985}.  For Mercury the MCNPX geometry consisted of a solid sphere with composition appropriate for the flyby geometry \citep{Peplowski2019}. A second MCNPX model of the MESSENGER spacecraft and detector was run to calculate the detectors' response to neutrons with different momenta. Finally, the formalism of \citet{Feldman1989} was used to calculate the detected count rate by analytically extending the surface flux to the flux at the spacecraft altitude and accounting for Doppler shifting of the detected flux due to the relative motion of the spacecraft and neutrons (Appendix~\ref{sec:Prop}).  This modeling builds on that from earlier nuclear spectroscopic studies \citep[e.g.,][]{Lawrence2010,Peplowski2015,Lawrence2013,Lawrence2017,Peplowski2019}.  A comparison of a subset of these models with the data taken during the Venus flyby is shown in Fig.~\ref{fig:cts}.

The final step in the modeling process was setting the absolute normalization of the models.  Normalization is required as our models of neutron production account for the GCR spectral shape but not for the absolute particle fluence. Ideally, the normalization would be set using the data at Venus to avoid the systematics associated with the measurements taken at Mercury.  However, for the 45 minutes of data that are available the statistics prevent this.  If the normalization is determined from the Venus data then the set of models with different parameters tend to overlap and although the shapes of the curves differ the statistics are not sufficient to distinguish between them. 

Separate normalization values were derived for each lifetime and were chosen to maximize the likelihood of the models given the data.  The GCR conditions during this time were almost identical to those during the Venus flyby and the spacecraft altitude and orientation were similar \citep{Peplowski2019}. More detail regarding this normalization, including a demonstration that an independent normalization derived at Venus is within 1\% of that derived at Mercury for models with a 900~s neutron lifetime, is given in \citet{Peplowski2019}. The statistical uncertainty in this normalization was included in the derived count-rate uncertainty at Venus. 

We validated our assumption that the absolute GCR flux is proportional to the normalization factor by examining how this factor changes with the solar modulation parameter $\Phi$.  Within the solar system GCRs are modulated by the Sun's magnetic activity. Thus, the modulation varies over the course of a solar cycle. $\Phi$ is used to characterize the shape of the modulated GCR spectrum and captures the temporal variation of the GCR flux \citep[e.g.,][]{Usoskin2011}. Using the flyby data and later orbital measurements the normalization factor can be seen to vary with $\Phi$ (Appendix~\ref{sec:SolarMod}).  This observation supports the claim that the normalization accounts for the absolute GCR flux and multiplying by these normalization factors incorporates this variation into the models.

\section{\label{sec:Results}Results}

\begin{figure}
 \includegraphics[width=\columnwidth]{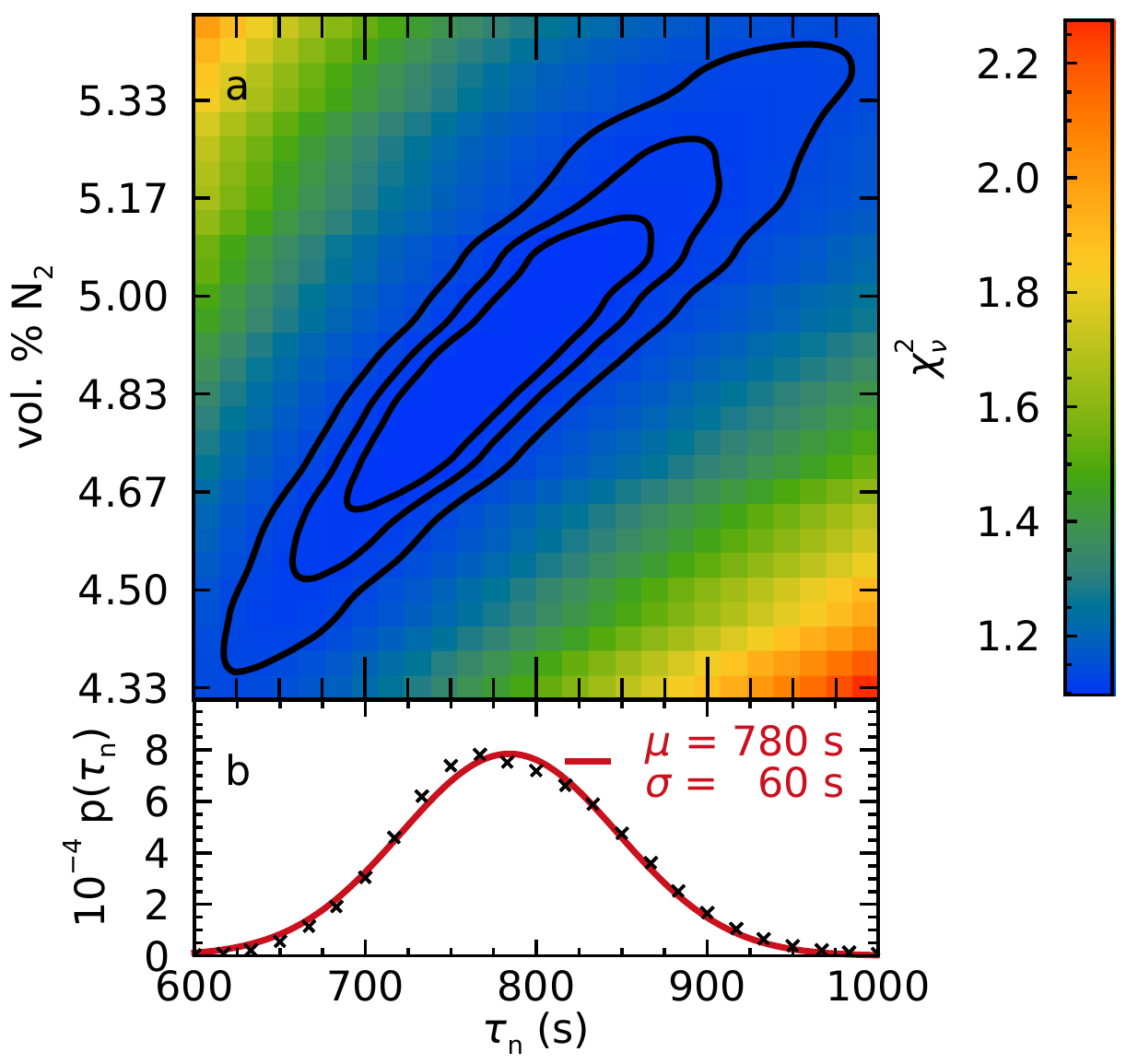}
\caption{\label{fig:CS_contours} (a) $\chi^2$ comparison of the models with differing $\tau_n$ values and N$_2$ abundances.  The contours show the 68, 95, and 99 \% confidence intervals, which were calculated assuming a Gaussian likelihood. (b) The probability distribution of $\tau_n$ using the likelihoods derived from (a).  The red curve is a Gaussian fit to the points, with the fit parameters shown.}
\end{figure}

The agreement between the data taken during the Venus flyby and models based on various atmospheric N$_2$ abundances and neutron lifetimes is shown in Fig.~\ref{fig:CS_contours}(a).  The minimum value of $\chi^2_\nu$ is consistent with unity. The single flyby of Venus, combined with the normalization determined at Mercury, provided sufficient data to constrain both parameters, as shown by the fact that the contours outlining the confidence intervals of the combined parameters are closed.  The ellipticity of these contours shows the degeneracy between the two parameters for our measurement. This is a result of the technique being sensitive primarily to the number of neutrons, which is reduced when decreasing $\tau_n$ or increasing N$_2$ abundance. A consequence of the degeneracy is that by combining prior constraints on $\tau_n$ with these MESSENGER data it is possible to make a more precise measurement of Venus' atmospheric N$_2$ content.  \Citet{Peplowski2019} used this approach to make the most precise measurement to date of the N$_2$ content of Venus' atmosphere at altitudes above 50~km.

Converting the $\chi^2$ values to likelihoods and integrating to marginalize out N$_2$ abundance enables the probability distribution of $\tau_n$ to be calculated in Fig.~\ref{fig:CS_contours}(b). This calculation implies $\tau_n=780\pm60_\textrm{stat}$~s, which is a 1.6$\sigma$ difference from the PDG value of $880.2\pm1.0$~s. The result demonstrates the feasibility of measuring $\tau_n$ using a space-based experiment.  Since the space-based method of constraining $\tau_n$ has a separate set of systematic uncertainties to the two existing classes of laboratory experiments, future space-based measurements with higher statistical precision than this current measurement at Venus may provide a route to make progress beyond the current disagreement between the bottle and beam results.

To discriminate between the two existing classes of measurement would require a precision of around 1\% or 9~s.  Obtaining a 1-$\sigma$ statistical uncertainty of 3~s, a factor of 20 smaller than the current estimate, would require a 400-fold increase in observation time, which translates to a 13~day observation period.  Of course for this to be practical requires the systematic errors to also be reduced beneath this level. 

%\subsection{Systematic uncertainties}
There are multiple systematic uncertainties that affect our estimate of $\tau_n$, summarized in Table~\ref{tab:sys}.  These can be split into two classes: those that affect only this particular implementation of the $\tau_n$-measurement and those that would affect any space-based methods using Venus as a neutron source.  The first class involves all of the systematics that result from taking observations at Mercury to set the normalization factor, which includes variation in the GCR flux between the Venus and Mercury flybys and uncertainty in Mercury's surface composition. In the second class are the effects of potential atmospheric compositional or thermal variation with latitude and time of day; species other than CO$_2$ and N$_2$ present in Venus' atmosphere; uncertainties in the modeled instrumental response function; and uncertainties in Monte Carlo particle transport modeling and the cross sections used. If a dedicated mission to perform a space-based $\tau_n$ measurement is to have uncertainties comparable in magnitude to existing laboratory measurements, then this second set and similar effects will need to be either mitigated or quantified and corrected. 

\begin{table}
\caption{\label{tab:sys}Summary of systematic uncertainties associated with the measurement of $\tau_n$ based on comparing models to data taken at Venus and Mercury.  Those that affect only this particular implementation of the neutron lifetime measurement are quantified.}
\begin{ruledtabular}
\begin{tabular}{lc}
Source of uncertainty &Uncertainty (s)\\
\hline
Mercury's surface composition & $\pm$70\\
% Mercury's surface inhomogeneity & $\pm$\later\\
Change in the GCR environment & $\pm$20\\\hline
\multicolumn{2}{l}{Instrument response function}\\
\multicolumn{2}{l}{Variation in Venus' atmosphere with time of day} \\
\multicolumn{2}{l}{Variation in Venus' atmosphere with latitude} \\
\multicolumn{2}{l}{Species other than CO$_2$ and N$_2$ in Venus' atmosphere} \\
\multicolumn{2}{l}{Uncertainties in the Monte Carlo modeling} 
\end{tabular}
\end{ruledtabular}
\end{table}

The largest identified systematic uncertainty is that on the model normalization associated with our imperfect knowledge of Mercury's surface composition, both in terms of its elemental makeup and how the distribution of elements varies across the planet's surface.  We estimate our uncertainty in Mercury's surface composition by considering the range in macroscopic neutron absorption cross sections $\Sigma_a$ measured across the planet's surface during MESSENGER's orbital mission at Mercury (4.5--${5.7 \times 10^{-3}}$~cm$^2$\,g$^{-1}$ \citep{Peplowski2015}). The absorption cross sections $\sigma_i$ are related to $\Sigma_a$ by
\begin{equation}
	\Sigma_a = N_A\sum_i\frac{\sigma_i w_i}{A_i},
\end{equation}
where $w_i$ is the weight fraction, $A_i$ the atomic mass of constituent $i$, and $N_A$ is Avogadro's number. With this definition, $\Sigma_a$ when multiplied by the material density is proportional to the probability per unit path length that a thermal neutron will be absorbed.

\begin{figure}[]
 \includegraphics[width=\columnwidth]{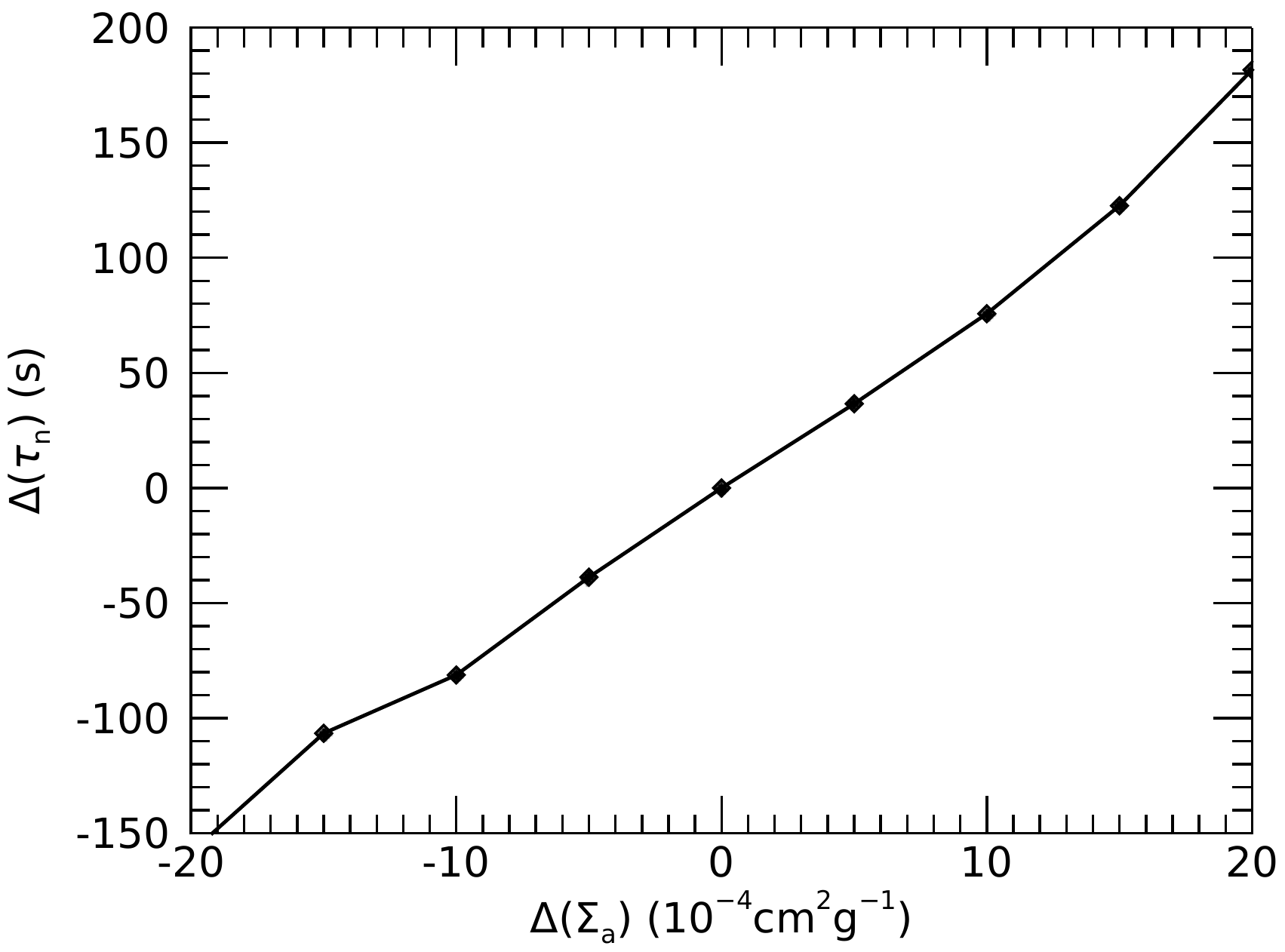}
\caption{\label{fig:S1} Change in $\tau_n$ inferred from the Venus flyby data when changing $\Sigma_a$ for the Mercury flyby.}
\end{figure}

To estimate the size of the systematic error introduced into our measurement of $\tau_n$ by this uncertainty in $\Sigma_a$, we produced a set of model observations based on different soil compositions. Fig.~\ref{fig:S1} shows the effect of changing $\Sigma_a$ on our estimate of $\tau_n$. The figure was created by varying the abundances of Cl and Fe by one multiplicative factor while changing the abundances of the other elements by a second factor, to ensure that the mass fractions sum to 1. As Cl and Fe were the elements in the reference composition with the largest neutron absorption cross sections these modifications had the effect of changing $\Sigma_a$.  Normalization factors were calculated using the model time series based on each of the new compositions.  Each of these normalization factors was then used to normalize the Venus models, which were then compared with the data.  This change in normalization caused our estimate of the lifetime to shift, which is shown in Fig.~\ref{fig:S1}. The full range of observed $\Sigma_a$ produced a change in our estimate of $\tau_n$ of $\pm 30$~s. Although $\Sigma_a$ is an important parameter in determining thermal neutron flux we found that model soil compositions with the same $\Sigma_a$ but different elemental distributions implied values of $\tau_n$ that differed by up to 60~s.  Taken in quadrature these errors imply a systematic uncertainty on the neutron lifetime associated with Mercury's surface composition of 70~s. 

In this analysis we assume that the GCR flux does not change between MESSENGER's observation at Venus and Mercury. We can place limits on this assumption using data taken by NASA's Advanced Composition Explorer.  Between the measurements the ACE-derived $\Phi$ values vary by $20\pm 40$~MV \citep{Peplowski2019}.  We can convert this to an expected change in the neutron count rate by considering how the NS detected triples count rate varies with ACE-derived $\Phi$.  A linear trend is shown in Fig.~\ref{fig:S2}b.  Consideration of this trend implies that a change in $\Phi$ of 20 MV between observation at Venus and Mercury yields a change in the normalization parameter of less than 3\%.  We can gauge the effect of such a change in normalization on $\tau_n$ by changing the value of the normalization used in the analysis.  Making this change produces a difference in $\tau_n$ of 20~s. 

It is clear from the preceding uncertainty estimates that using data taken at Mercury to provide the model normalization introduces a model-dependence that is absent from the original, optimized form of a space-based measurement. This compromise is a consequence of the fact that this mission was not designed with the goal of measuring $\tau_n$ but of answering several other questions in planetary science.  However, the success of a measurement using this suboptimal dataset demonstrates the feasibility of measuring $\tau_n$ from space and provides an initial step on the path to flying an optimized mission.  These systematics could be avoided if more data were taken at Venus, because improving the statistics of the Venus measurements would enable that data set to be used alone. The measurement of $\tau_n$ would then not be set by the mean count rate in the models but only by how the detected neutron count rate changes with altitude.

The remaining systematics would affect any experiment conducted at Venus (or another planet with a thick atmosphere such as Earth). We expect these unquantified systematics to be smaller than those discussed above.  However, we leave their detailed exploration for a study focusing on the practicalities of a future mission rather than this proof-of-principle study. For this particular, MESSENGER-based measurement the total systematic uncertainty is 70~s.

\section{\label{sec:Conclusions}Conclusions}

Using data taken by MESSENGER's Neutron Spectrometer during its flybys of Venus and Mercury we found ${\tau_n=780\pm60_\textrm{stat}\pm70_\textrm{syst}}$~s.  This result establishes the feasibility of making a measurement of $\tau_n$ from space. The statistical uncertainties are large due to the short duration of the flybys (totaling 70 minutes with altitude below $10^4$~km) and subsequent small amount of data taken, which is a consequence of the mission not being planned with this measurement in mind. The systematic errors are similarly large; however, the worst of these could be avoided with a longer duration experiment using observations taken only at Venus thus avoiding the systematics associated with uncertainties in Mercury's surface composition. The reduction of smaller-magnitude systematics to the 1~s level required to potentially resolve the neutron lifetime anomaly requires a detailed mission design study that builds on the result of this paper. 

\begin{acknowledgments}
We gratefully acknowledge the support of the U.S. Department of Energy Office of Nuclear Physics (Grant No. DE-SC0019343). VRE was supported by the Science and Technology Facilities Council (STFC) grant ST/P000541/1. JAK acknowledges support from STFC grant ST/N50404X/1 and the ICC PhD Scholarships Fund. 
\end{acknowledgments}

\appendix
\section{\label{sec:Prop}Neutron Propagation}
The propagation of neutrons in a spherically symmetric gravitational field is described in \citet{Feldman1989} and our implementation follows that work. The neutron flux at altitude $R$ can be expressed in terms of the surface flux $\phi(K,\mu)$ as
\begin{equation}
	\phi_R(K_R,\mu_R) = \sqrt{\frac{K_R}{K}}\phi(K,\mu)\exp\left(\frac{\Delta t_R}{\tau_n}\right)
\end{equation}
where $K$ is the kinetic energy of the neutron at emission, $K_R=K - V(R-R_0)$ is the kinetic energy at $R$, with $V = \frac{GMm}{R_0}$, $M$ the mass of the planet, $m$ the neutron mass, $\mu$ the cosine of the angle of emission with respect to the local zenith $\theta$, $\mu_R = \sqrt{1-(R_0/R)^2(K/K_R)(1-\mu^2)}$, and $\Delta t_R$ is the time for transit for a neutron travelling from the surface to an altitude $R$ \citep{Feldman1989}.  The expression for $\Delta t_R$ in \citet{Feldman1989} contains an extraneous minus sign on $\mu$.  For reference, the corrected expression is given here
\begin{multline}
	\Delta t_R = \frac{R_0(m/2V)^{1/2}}{2(1-K/V)^{3/2}}\left(2\mu\left(1-\frac{K}{V}\right)^{1/2}\left(\frac{K}{V}\right)^{1/2}\left(1 - \left|\frac{\tan{\theta}}{\tan{\theta_R}}\right|\right) \right. \\ \left. + \sin^{-1}\left(\frac{B}{(A^2+B^2)^{1/2}} \right) + \sin^{-1}\left(\frac{1 - 2K_R/V_R}{(A^2+B^2)^{1/2}} \right) \right)
\end{multline}
where 
\begin{equation}
	A = \sqrt{4\left(\frac{K}{V}\right)\left(1 - \frac{K}{R}\right)\mu^2}, \quad B = \left(\frac{2K}{V} - 1\right) \quad \text{for}\quad \frac{K}{V} < 1.
\end{equation}

\begin{figure}
 \includegraphics[width=\columnwidth]{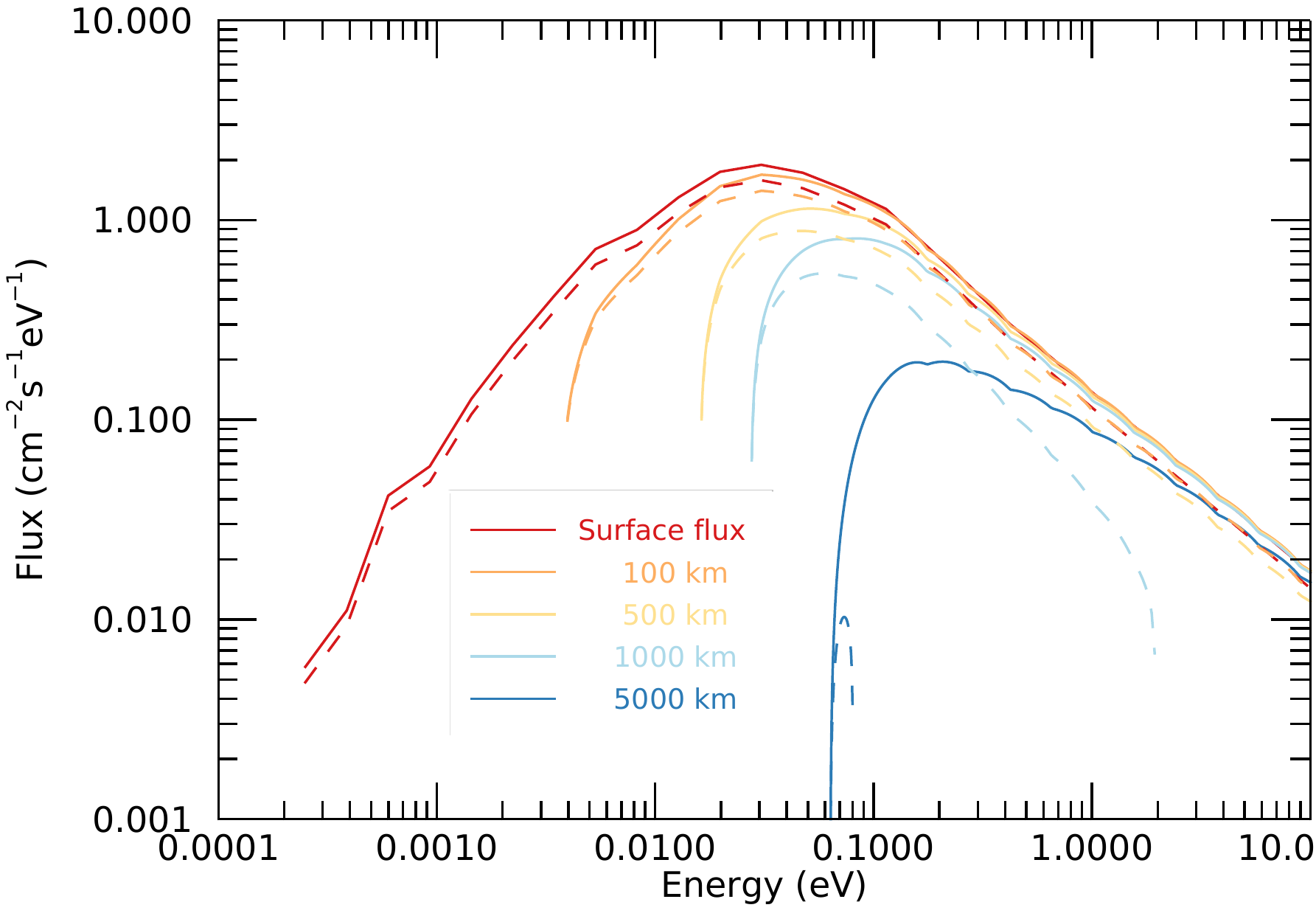}
\caption{\label{fig:SFlux} The neutron flux at several altitudes above Mercury. The solid curves show neutrons detected with $\mu_R=1$ and dashed curves $\mu_R=0.7$.}
\end{figure}

Fig.~\ref{fig:SFlux} shows the simulated flux at Mercury's surface along with the fluxes at a range of altitudes above the surface determined by the equations above.  For neutrons from a planetary surface the emitted flux is approximately proportional to the square root of the normal to the emission angle $\mu$ \citep{Lawrence2006}.  The two most obvious features in the plot are the low energy cutoff due to conversion of kinetic to gravitational potential energy as neutrons rise out of the planet's potential well and, for the neutrons received at an angle to the local zenith, the high-energy, high-altitude cutoff, which is due to the lower curvature of higher-energy neutrons and the decreasing solid angle occupied by the planet with increasing altitude. 

The transformation from the flux at altitude to that observed by the spacecraft requires a Galilean transformation.  The mean velocity during the Venus flyby was 10.6~km\,s$^{-1}$, which is the speed of a neutron with a kinetic energy of 0.6~eV (the maximum velocity was 13.5~km\,s$^{-1}$, which corresponds to a 0.95~eV neutron).  At Mercury the spacecraft had a mean velocity of 6~km\,s$^{-1}$ (0.2~eV).  

\section{\label{sec:SolarMod}Variation of Model Normalization with Solar Modulation Parameter}

\begin{figure}
 \includegraphics[width=\columnwidth]{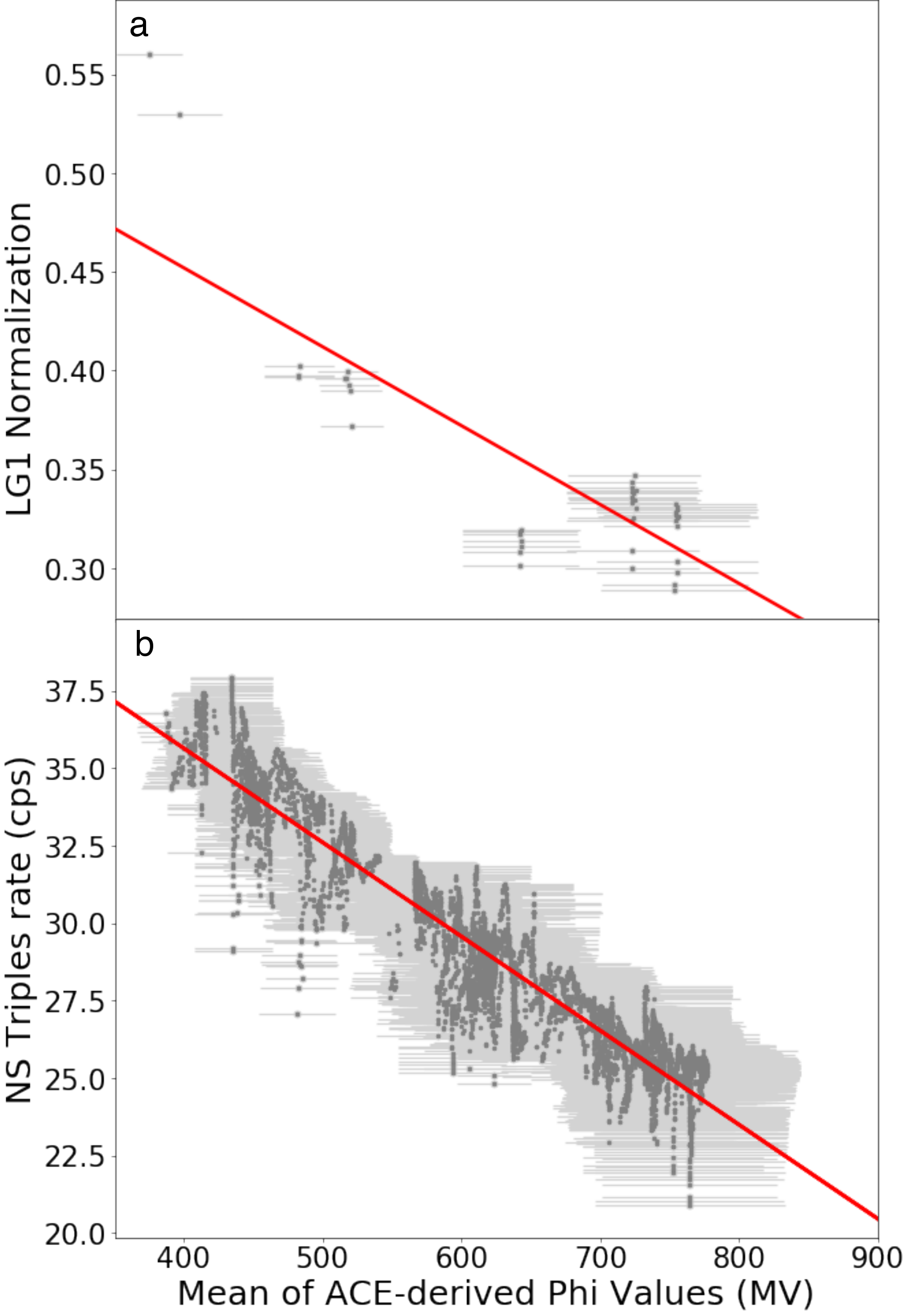}
\caption{\label{fig:S2} (a) Change in data-to-model normalization factor with solar ACE-derived modulation parameter. (b) Change in the MESSENGER NS triples count (a proxy for GCR flux) with ACE-derived solar modulation parameter. The red curves in both panels are linear fits to the points.}
\end{figure}

The GCR spectrum can be characterized by a solar modulation parameter $\Phi$, which describes the intensity and spectral shape of GCRs as they respond to changes in the solar magnetic field. Higher values of $\Phi$ correspond to fewer GCR protons in the inner solar system.  To demonstrate the relationship between the model normalization factor and GCR flux we calculated normalizations for a range of dates with different $\Phi$ values taken during the flybys (the two low-$\Phi$ points in Fig.~\ref{fig:S2}a) and the later orbital data.  Fig.~\ref{fig:S2}a shows that as $\Phi$, measured by NASA's Advanced Composition Explorer (ACE), decreases the required factor to normalize the models to the measured data increases, which is as expected if this factor accounts for changing absolute neutron flux.

MESSENGER's NS had a triple coincidence mode whereby GCRs with energy above 120 MeV are registered when all three NS scintillators are triggered in coincidence.  This `triples' count rate has been shown to be a good proxy for local GCR flux \citep{Lawrence2013}. Thus, comparison of the ACE-derived $\Phi$ and triples provides an indication of how much the neutron count rate should be expected to change with $\Phi$.  Fig.~\ref{fig:S2}b shows the trend in triples counts with $\Phi$. As the ACE data are reported approximately monthly and the triples once per orbit, or several measurements per day, the NS triples measurement captures short-term variability not reported by ACE, which is apparent in the plot.

\end{document}